\documentclass[sigconf]{acmart}

\def\BibTeX{{\rm B\kern-.05em{\sc i\kern-.025em b}\kern-.08emT\kern-.1667em\lower.7ex\hbox{E}\kern-.125emX}}
    
\copyrightyear{2019} 
\acmYear{2019} 
\acmConference[WI '19]{IEEE/WIC/ACM International Conference on Web Intelligence}{October 14--17, 2019}{Thessaloniki, Greece}
\acmBooktitle{IEEE/WIC/ACM International Conference on Web Intelligence (WI '19), October 14--17, 2019, Thessaloniki, Greece}
\acmPrice{15.00}
\acmDOI{10.1145/3350546.3352509}
\acmISBN{978-1-4503-6934-3/19/10}
\usepackage{color}
\usepackage{amsmath}
\usepackage{cases}
\usepackage{amsfonts}
\usepackage[ruled]{algorithm2e}

\makeatletter
\newcommand{\figcaption}[1]{\def\@captype{figure}\caption{#1}}
\newcommand{\tblcaption}[1]{\def\@captype{table}\caption{#1}}
\makeatother

\usepackage{subfigure}

\usepackage{enumerate}
\begin{document}

\title[Immediate Personalized News Recommendations]{Algorithms and System Architecture for Immediate Personalized News Recommendations}

\author{Takeshi Yoneda}
\affiliation{%
  \institution{Gunosy Inc.}
  \city{Tokyo}
  \country{Japan}
}
\email{takeshi.yoneda@gunosy.com}

\author{Shunsuke Kozawa}
\affiliation{%
  \institution{Gunosy Inc.}
  \city{Tokyo}
  \country{Japan}
}
\email{shunsuke.kozawa@gunosy.com}

\author{Keisuke Osone}
\affiliation{%
  \institution{Gunosy Inc.}
  \city{Japan}
  \country{Tokyo}}
\email{keisuke.osone@gunosy.com}

\author{Yukinori Koide}
\affiliation{%
  \institution{Gunosy Inc.}
  \city{Japan}
  \country{Tokyo}}
\email{yukinori.koide@gunosy.com}

\author{Yosuke Abe}
\affiliation{%
  \institution{Gunosy Inc.}
  \city{Japan}
  \country{Tokyo}}
\email{yosuke.abe@gunosy.com}

\author{Yoshifumi Seki}
\affiliation{%
  \institution{Gunosy Inc.}
  \city{Japan}
  \country{Tokyo}}
\email{yoshifumi.seki@gunosy.com}

\renewcommand{\shortauthors}{Yoneda et al.}

%
\begin{abstract}

Personalization plays an important role in many services, just as news does. 
Many studies have examined news personalization algorithms, but few have considered practical environments. 
This paper provides algorithms and system architecture for generating immediate personalized news in a practical environment.
Immediacy means changes in news trends and user interests are reflected in recommended news lists quickly. 
Since news trends and user interests rapidly change, immediacy is critical in news personalization applications.
We develop algorithms and system architecture to realize immediacy.
Our algorithms are based on collaborative filtering of user clusters and evaluate news articles using click-through rate and decay scores based on the time elapsed since the user's last access. 
Existing studies have not fully discussed system architecture, so a major contribution of this paper is that we demonstrate a system architecture and realize our algorithms and a configuration example implemented on top of Amazon Web Services. 
We evaluate the proposed method both offline and online. 
The offline experiments are conducted through a real-world dataset from a commercial news delivery service, and online experiments are conducted via A/B testing on production environments.
We confirm the effectiveness of our proposed method and also that our system architecture can operate in large-scale production environments.
\end{abstract}

%
%

%

%
\maketitle

\section{Introduction}
Continual increases in the number of news articles available online make it difficult for users to select news articles that correspond to their interests.
Major online news distribution services, such as Google News\footnote{https://news.google.com/} and Yahoo!,\footnote{https://www.yahoo.com/} use various methods to offer selected articles that may interest users from a large body of collected news articles.

We propose a novel personalized news recommendation system. 
Personalization plays an important role in various web service applications, such as e-commerce, streaming, and news. 
Since it is important to construct an appropriate recommendation algorithm for each service, many studies have focused on the needs of various online services. 
News differs from e-commerce and streaming, as item lifetimes are short and trends and user interests change rapidly \cite{ozgobek2014survey}. 
These differences create challenges in designing news personalization algorithms

Our system focuses on immediacy. 
Immediacy means changes in news trends and user interests are reflected in recommended news lists quickly. 
We expect that an immediate news recommendation system would be able to quickly recommend high-value news articles to users. 
Although some studies have proposed algorithms that capture the change of user interest or popularity, no studies have provided solutions to reflect and update these change immediately.

Our news recommendation system includes the following features: 
\begin{itemize}
    \item cluster-based collaborative filtering (CF) using click-through rate (CTR); 
    \item incorporating a user time decay function (UTDF); and
    \item system architecture that is able to update and reflect changes immediately. 
\end{itemize}

We adopt cluster-based CF scoring according to the CTR to reflect current popularity.
CF is a powerful algorithm used in many services \cite{Su}.
As traditional CF uses numbers of actions (e.g., conversions and clicks) to produce scores, new items are undervalued and old items are overvalued.
Fresh news articles should be treated as highly valued, but when click numbers are used for scoring news articles, these articles are not valued as highly as should be; therefore, we use CTR to simplify the valuation of such articles. 
Using CTR makes it possible to give news articles a high score if they are clicked with a high probability (high CTR), even if the number of times displayed is small.
In addition, if the number of clicks is high but popularity is lost as a result of staleness, the CTR will decline; therefore, it is possible to reduce the value of old news articles. 

To evaluate the freshness correctly in a personalized way, we calculate how new a news article is for each user using the user's last access time. 
Then, the personalized recommendation scores for each user are decayed according to that freshness.
For example, users who access a news service many times a day only consider articles published in the last few hours to be fresh, while users who access such services once a day consider articles fresh even if they were published half a day ago.
Given that, it is plausible to adopt a user time decay function (UDTF) that decays scores using the time elapsed since the user's last access.

Those features should be reflected and updated in real time, so we designed and implemented a system architecture that can do so.
Although existing studies have not fully discussed an architecture, designing a practical and concrete system architecture is important.
We show real-world working system architecture to realize our algorithms and the configuration example implemented on top of Amazon Web Service (AWS).

Our contributions are summarized as follows: 
\begin{itemize}
    \item a news recommendation system focusing on immediacy is proposed; 
    \item the system architecture and a configuration example of our algorithms are shown; and 
    \item our system's efficiency is demonstrated through both offline and online experiments. 
\end{itemize}

The rest of this paper is as follows. 
In section \ref{section:algorithms}, we describe the proposed scoring function based on the CF of user clusters, which is a combination of the user modeling, the news article evaluation using CTR, and a UTDF.
In section \ref{SystemArchitecture}, we show our system architecture and describe how it is implemented on top of AWS from software engineering points of views.
In section \ref{Experiments}, our online/offline experiments' methodology and results are shown.
In section \ref{section:related_work}, related works are shown along with their relevance to this study.
Section \ref{section:conclusion} presents our conclusions.

\section{Algorithms}\label{section:algorithms}

In this section, we describe algorithms used in our proposed method.
The main concept addressed by our method is immediacy.
Moreover, since we aim for these algorithms be deployed in a million-scale production environment, they must be scalable.
To capture changes in user interest, our user modeling algorithm is able to update the user's feature in real time, and its system architecture is scalable. 
Since the value of news articles declines as their freshness diminishes, we adopt a UTDF that decays scores alongside the user's last access time. 
In addition, to reflect these captured changes and produce a recommendation, our scoring algorithm formulates a matrix operation with a scalable system architecture. Therefore, our recommendation system is able to generate an immediate recommendation.

\subsection{User Modeling}\label{subsec:user-model}

Our proposed method uses click action to model user behavior. 
First, we define the vector representation of a news article; then, we define the vector representation of a user through the article's vector representation based on the user's click actions.
By $a$, we denote a news article.
Let $W_a := \{w_i \}_i \subset \mathbb{R}^d$ represent the set of words contained in $a$. 
Here, we consider every word $w_i$ to be a $d$-dimensional dense vector obtained from the word2vec model \cite{word2vec} that we trained on sentences from millions of past news articles.\footnote{More precisely, we trained a Continuous Bag of Words (CBoW) model with a dimension of 300 and a window size of 3.}

Then, we define news article $a$'s vector representation as follows:
\begin{eqnarray*}
a := \dfrac{\sum_{w_i \in W_a} \it{idf}(w_i) w_i}{\left\|\sum_{w_i \in W_a} \it{idf}(w_i) w_i \right\|} \in \mathbb{R}^d,
\end{eqnarray*}
where $\it{idf}(w_i)$ is the inverse document frequency calculated from past news articles.
According to our prior experiment, we use the weighted average of the vectors rather than their simple average.

Now we are ready to define the vector representation of users.
Let $u$ be a user and $A_u$ be the set of the last $N$ news articles clicked by $u$. Then, let us define the vector representation of $u$ as follows:
\begin{eqnarray}\label{eq:user_model}
u := \dfrac{\sum_{a \in A_u} a}{\| \sum_{a \in A_u} a \|} \in \mathbb{R}^d,
\end{eqnarray}
which is the average of $A_u$ news articles' vectors. 

The advantage of this approach is that $u$ can be updated as soon as they click on a news article.
New users' vector representations can be obtained immediately after they click on only one news article.
Additional details on the process of updating are described in Section \ref{SystemArchitecture}.

\subsection{Scoring Algorithm}

We adopt a CF-based algorithm, which recommends items that are popular among similar users. 
CF is a major personalization algorithm. 
Although \cite{yahoo} suggests that a content recommendation system should be used in a news recommendation system, in its system, manually selected major news articles are distributed separately from the recommendation system's results.
We aim to construct a news service using only a personalized algorithm; therefore, our personalized algorithms should be able to deliver major news articles. 
The conditions of major news vary but surely include popularity among users, which is why we adopt an algorithm based on CF that can deliver major news articles to interested users.

In addition, although many news recommendation systems applying CF use click numbers to score items, in this study, we use CTR to more easily score fresh news articles.
One of the challenges of CF is that it is difficult to evaluate new items properly. 
Fresh news articles are valuable, but they are difficult to be valuated through click numbers. 
Using CTR makes it possible to value novel news articles highly if they are clicked with a high probability, even if the display number is small.
In addition, if an article's number of clicks is high but its popularity is low as a result of staleness, its CTR will decline; therefore, it is possible to lower the value of older news articles. 
News articles that have never been displayed are not scored, even using CTR; this problem can be solved by displaying such articles a certain number of times, for example, with a bandit algorithm or CTR-prediction models. These are beyond the scope of this paper.

Let $A$ be the set of news articles and $U$ be the set of users.
The process of calculating a score for a paired user $u \in U$ and news article $a \in A$ is as follows:
\begin{description}
    \item[Step1] Execute unsupervised clustering of users.
    \item[Step2] Calculate the recommendation score using the cluster-based CF algorithm.
    \item[Step3] Decay the score of $a$ using the user's last access time and the article's publication time.
\end{description}
Step 1 is executed beforehand as a batch process, and steps 2 and 3 are executed in response to the user's request.

The user is vectorized by equation (\ref{eq:user_model}) and is classified into $K$ clusters using k-means clustering through step 1. 
Let $C = \{c_1, ..., c_K\}$ be the set of calculated centroid vectors of each cluster and $U_{c_{i}}$ be the set of users belonging to the cluster whose centroid vector is $c_i$.

As for steps 2 and 3, the scoring function based on cluster-based CF is calculated as follows:
\begin{eqnarray}\label{eq:score}
    score(u, a) = t(u, a) \sum_{c_i \in C} w(u, c_i) \times {\rm CTR}(U_{c_i}, a).
\end{eqnarray}
Here, $t(u, a)$ represents decaying score by time and is described in Section \ref{subsec:timedecay}. Other terms are explained as follows.
${\rm CTR}(U_{c_i}, a)$ is the observed CTR in $U_{c_i}$ of article $a$, and the $w: \mathbb{R}^d \times \mathbb{R}^d \rightarrow \mathbb{R}_{\geq 0}$ can be any function with the following property:
the shorter the distance between $u$ and $c_i$ is, the larger $w(u, c_i)$ is. The choice of $w$ is described in section \ref{Experiments}.
In the scoring function, the weighted CTR is treated as the score according to the distance between the user vector and the centroid vector of the cluster. 
Since the user vector is updated in real time with user clicks, the score is also updated in real time.

This scoring function is highly consistent with that in \cite{google}, although the clustering algorithm, the distance to the cluster, and the evaluation of the article are different.
Since this scoring function is calculated as a matrix operation, it is possible to operate it on demand at high speeds even if the number of candidate news articles is significant.

\subsection{Time Decay Function}\label{subsec:timedecay}

The freshness of news articles is an important value. Here, we describe $t(u, a)$ in equation (\ref{eq:score}).
Thus, we adopt a function that decays the score according to the time that an article was published, which we call the time decay function (TDF).
TDF is defined as follows:
\begin{align}
\label{tdf}
TDF(u, a) &:=   \left\{
\begin{array}{ll}
1 & (T_{\it tdf} < \Delta_{t_a}) \\\exp\left( \dfrac{ - ( T_{\it tdf} - \Delta_{t_a})^2 }{2 \sigma}  \right) & (T_{\it tdf} \geq  \Delta_{t_a}),
\end{array}
\right.
\end{align}
\begin{align*}
\Delta_{t_a} := t_{\rm now} - t_a.
\end{align*}
where $t_{\it now}$ is the current time, $t_a$ is the publication time of news article $a$, and $\Delta_{t_a}$ is the elapsed time since the publication of news article $a$.
$T_{\it tdf}$ is the threshold of elapsed time, and $\sigma$ is a constant that determines the scale of decays.
In this way, TDF exponentially decays the score according to the elapsed time since the publication of the news article.
Using TDF as $t(u, a)$ in equation (\ref{eq:score}), the resulted recommendation system is able to valuate fresh news articles more highly than old ones. 

However, the freshness of new articles differs by user; the freshness of an article for users who accessed the service in the past hour is different from the same article's freshness for users who accessed the service only in the past day. 
Therefore, we propose a UTDF that decays the score using the user's last access time. 
We define UTDF as follows: 
\begin{align}
\label{utdf}
UTDF(u, a) &:=   \left\{
\begin{array}{ll}
1 & (T_{\it utdf} < \Delta_{t_{u,a}}) \\\exp\left( \dfrac{ - ( T_{\it  utdf} - \Delta_{t_{u,a}})^2 }{2 \sigma}  \right) & (T_{\it utdf} \geq  \Delta_{t_{u,a}}),
\end{array}
\right.
\end{align}
\begin{align*}
\Delta_{t_{u, a}} &:= t_{u} - t_a, 
\end{align*}
where $t_u$ is the last access time of user $u$, $\delta_{t_{u,a}}$ is the elapsed time from the last access time of user $u$ to the publication time of article $a$, and $T_{\it utdf}$ is the threshold of elapsed time.
As as a result, UTDF is able to decay the score according to the user's last access time in the personalized way.
The effects of using a UTDF are discussed in section \ref{Experiments}.

\section{System Architecture}\label{SystemArchitecture}
\begin{figure*}[tp]
  \centering
  \includegraphics[width=0.99\linewidth]{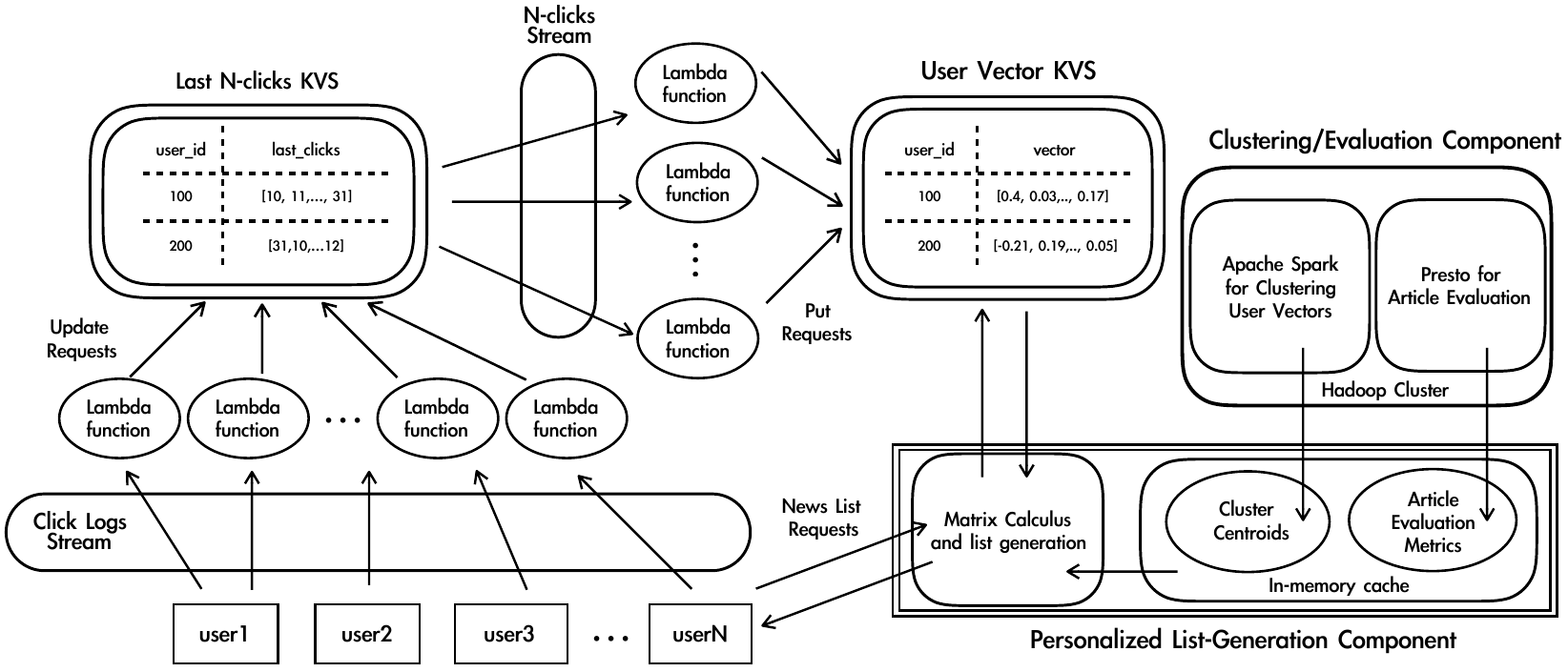}
    \caption{System Architecture Overview: Our system consists of three components. The user modeling component extracts a log of a user's last N-clicks from the log streaming and calculates the user vectors. The clustering and evaluating component assigns each user to a cluster and aggregates the application logs to calculate the ${\rm CTR}(c_i, a)$ of every candidate news article in each cluster. The personalized list generation component generates personalized news article lists in response to user requests in real time.}
    \label{fig:system_architecture}
\end{figure*}

As shown in Figure \ref{fig:system_architecture}, our system consists of the following three components:
\begin{itemize}
\item {\bf User Modeling}: calculate equation (\ref{eq:user_model}) and keep the user model up to date.
\item {\bf Clustering and Evaluation}: cluster the user models, and calculate ${\rm CTR}(c_i, a)$. 
\item {\bf Personalized List Generation}: generate personalized news article lists instantly in response to requests by users at scale.
\end{itemize}

Below, we describe these components in detail.
All components are running on top of AWS.

\subsection{User Modeling Component}

The process for the user modeling component is as follows:
\begin{enumerate}[(i)]
    \item Receive a user click log from the log stream.
    \item Add the received click data to the list of user's click history; if the length reaches the limit, trim.
    \item Generate the user model from user's click history.
\end{enumerate}
The service's action log is the streaming data.
When the user clicks on a news article, a click log is put in the stream.
The list of user's click history is saved in Key Value Store (KVS) as List Type with the user-id as the Key.
When the click log is received, the clicked article's ID is pushed to the list of the Key corresponding to the clicked user's ID and is trimmed if the list size exceeds the specified number. 
After the list of click history is updated, the user model is recalculated and stored according to equation (\ref{eq:user_model}).
Through the above process, the user model is always kept up to date.

We constructed the above process in the serverless manner using some AWS services.
Amazon Kinesis Stream was used for log streaming, Amazon DynamoDB for KVS, and Amazon Lambda for calculating user models and pushing and trimming click data.
Since these services can be operated as event-driven and are scaled automatically according to the workload, this pipeline is applicable to large-scale news recommendation services.

\subsection{Clustering and Evaluation Component}
This component executes k-means clustering and calculates the CTR of news articles in each cluster ${\rm CTR}(U_{c_i}, a)$.
This is necessary to calculate the scoring function defined in equation (\ref{eq:score}).
For scalability purpose, we have adopted Apache Spark\footnote{https://spark.apache.org} and Presto.\footnote{https://prestodb.io}
These software run on Amazon EMR and are suitable for distributed computation.\footnote{https://aws.amazon.com/emr}

The k-means clustering algorithm is run using Apache Spark.
Since we have adopted parallelized k-means++ \cite{kmeans}, we can scale huge and high-dimensional datasets.
Clustering is conducted every day for all users who have vector, and we store each user's assigned cluster for ${\rm CTR}(U_{c_i}, a)$ calculation using Presto. In addition, each cluster's centroid vector are stored in DynamoDB for the Personalized List Generation Component described in Section \ref{subsec:personalized_list_geneartion}.

The ${{\rm CTR}(U_{c_i}, a)}$ is calculated using Presto -- one of the distributed query engines that executes the calculation efficiently and at high speeds.
These features are helpful since the amount of the whole log data is huge and the computational cost is high to calculate ${\rm CTR}(c_i, a)$.
We calculate and update ${{\rm CTR}(U_{c_i}, a)}$ every 10 minutes.

\subsection{Personalized List Generation Component}\label{subsec:personalized_list_geneartion}
This component generates personalized news article lists instantly in response to user requests.
These servers load cluster centroid vectors $c_1, \dots, c_K$ and ${\rm CTR}(U_{c_i}, a)$ into their in-memory cache asynchronously with user requests.
When a user sends a personalized news list request to the server, the service initiates
\begin{enumerate}[(i)]
    \item gets the user's vector for the requested user from KVS;
    \item calculates $\{w(u, c_i)\}_{i}$, distances between the user $u$ and all cluster centroid vectors;
    \item calculates $ \{ score(u, a) \}_{a \in A}$ defined in equation (\ref{eq:score}), the score between user $u$ and all candidate news articles $a \in A$; and
    \item sorts the candidate news articles by the calculated scores and returns the top $M$ news articles as a personalized news list. 
\end{enumerate}
The average response time is less than 25 milliseconds, which is sufficient to deploy these components in million-scale user environments and achieve user satisfaction.

Since the list generation is performed only when there is a request from a user, it is not necessary to preprocess for those who will not make a request.
This is computationally efficient compared to the common cases where it is necessary to calculate and prepare for all users, including the ones who never send requests.

\section{Experiment and Evaluation}\label{Experiments}

We conducted offline and online experiments to evaluate the proposed method. 
The dataset for the offline experiment was sampled from user behavior log in a popular Japanese mobile news application called NewsPass.\footnote{https://newspass.jp/} 
The online experiment was conducted through this application via an A/B test. 
Through the offline experiment, we confirmed the effectiveness of the proposed user modeling and cluster-based CF, 
and through the online experiment, we confirmed the effectiveness of UTDF.
Moreover, we verified that our system architecture works in the million-scale production environment.

As a choice for $w: \mathbb{R}^d \times \mathbb{R}^d \rightarrow \mathbb{R}_{\geq 0}$ in our proposed scoring function in equation (\ref{eq:score}), we used the following simple function throughout both in offline and online experiments: 
\begin{eqnarray*}
w(u, v):= \dfrac{1}{ \|u - v\|^{10}}.
\end{eqnarray*}
This was chosen based on our prior experiments through multileaving methods.
\subsection{Offline Experiment: Dataset}\label{subsec:overview_offline}

This experiment was conducted using user action logs from NewsPass, which provide categorized news articles lists as specific tabs.
The target period was a certain week in 2017, and the target users were sampled from a user group narrowed down by specific conditions.
The first tab that opens when a user launches the application is the {\it topic tab}; it is the most actively used tab.
In this experiment, we only used logs from the topic tab. 

The following datasets were used in the offline experiment:
\begin{itemize}
    \item recommended candidate news articles;
    \item for evaluation, news articles from the topic tab clicked by users;
    \item click history for constructing user vectors; and
    \item the CTR of candidate news articles for calculating the score.
\end{itemize}

To recommend candidate news articles, we displayed news articles to users in the topic tab. 
To evaluate our recommendation methods, we used click logs from the news articles in the topic tab. 
In this experiment, each recommendation method was used to sort candidate news articles. 
When news articles that users clicked on occupied high positions in the sorted news article list, we cloud claim that that method worked as it should.
To construct a user vector, we used all the click logs. 
The CTR of each cluster was used in the scoring function as described in equation \ref{eq:score}.
Since the clustering result varies depending on the method and duration of the experiment, to calculate the CTR of each cluster in each period, the datasets included display and click logs.

This dataset was constructed according to the action logs of target users. 
We narrowed down target users through the following two criteria:
\begin{itemize}
    \item the user clicked on one or more news articles from the topic tab; and
    \item the user had a user vector within the dataset.
\end{itemize}
The number of users fulfilling these conditions cannot be disclosed due to a business issue. 
Of this number, 30,000 users were randomly sampled as target users.

\subsection{Offline Experiment: Settings}

In the offline experiment, we conducted two experiments that prepared different candidate articles. 
The experiments were designed to evaluate a recommendation list hourly, so candidate articles were prepared every hour. 
We used articles displayed in the topic tab as recommended candidate articles.

The first experiment presented all articles available each hour for all users as candidate articles, and this experiment was called the {\it all experiment}.
Although this experiment created an environment similar to the actual environment that evaluates many articles, it was difficult to evaluate  
because articles that were not displayed to the user were also recommended.

In the second experiment, different candidate articles were prepared for each user, and this experiment was called the {\it user experiment}.
In this experiment, articles were displayed to the user every hour as candidate articles. 
Since this experiment was evaluated only through displayed articles, we expected that the articles' valuation result would be similar to users' satisfaction with them. 
However, the articles displayed to the user varied according to the order of the articles in the existing system. 
The existing system sorts articles by the CTR of a user's demographic data (e.g., gender or age), and it operates under several rules. 
 Therefore, the candidate news articles consisted of many high CTR news articles, which led to an overestimation of the performance of some methods.

In offline experiments, the TDF and the UTDF, described in section\ref{subsec:timedecay}, freeze as $t(v, a) = 1$.
Offline experiments confirmed the effectiveness of cluster-based CF using CTR, and the effectiveness of the UTDF was confirmed in the online experiments. 
The reasons for this are as follows:
\begin{itemize}
    \item the computational costs of the UTDF in the offline experiments were too high; and
    \item we expected that the UTDF would be strongly effective in the online experiment. 
\end{itemize}

{\bf Methods for Comparison.} 
In the offline experiment, we compared the proposed method with a simple content-based filtering method 
and related CF-based methods (MinHash and pLSI) \cite{google}.
Using the simple content-based filtering method, we calculated the inner product of a user model and an article vector as the score. 
This method is similar to the baseline method discussed in \cite{yahoo}.
In this experiment, we implemented non-negative matrix factorization (NMF) \cite{lee1999} for pLSI because, in \cite{ding2008}, pLSI had a common objective function with NMF and the two were shown to be equivalent.

The difference between the proposed method and the related CF-based methods is their clustering and news article scoring function. 
For clustering, the proposed method uses k-means+w2v clustering, while the related methods use MinHash and NMF. 
In regard to how articles are scored, the proposed method uses CTR, while the related methods use click numbers. 
Therefore, to examine the validity of the proposed method, we must compare all combinations of the clustering algorithms (k-means+w2v, MinHash, and NMF) and the ways of scoring articles (CTR and number of clicks).

{\bf Evaluation Metrics.}
We evaluated the method by the positions of the news articles that the user clicked in the sorted list, 
so we used mean average precision (MAP) and normalized discounted cumulative gain (NDCG) as evaluation metrics. 
The {\it all experiment} had many candidate articles; thus, we adopted MAP@10 and NDCG@10. 
In the {\it user experiment}, only articles displayed to the user were candidate articles; thus, we were able to calculate MAP and NDCG for the entire list.
In order to evaluate immediacy, these experiments were conducted on separate datasets each hour, and these evaluation metrics were calculated for each segment.

\subsection{Offline Experiment Results}
\begin{table}[tp]
\caption{Both Offline Experiment Results: We compared methods using MAP and NDCG. The best performance is highlighted in bold; the proposed method performed the best.}
\label{tbl:result_all}
\begin{center}
\small
\begin{tabular}{|l|c|c|c|c|}
\hline
                            & \multicolumn{2}{c|}{{\it all experiment}} & \multicolumn{2}{c|}{{\it user experiment}} \\ \hline
                            & MAP@10           & NDCG@10          & MAP               & NDCG             \\ \hline
simple content-based        & 0.043            & 0.056            & 0.436             & 0.468            \\ \hline
MinHash (Clicks)            & 0.077            & 0.094            & 0.419             & 0.452            \\ \hline
MinHash (CTR)               & 0.080            & 0.090            & 0.324             & 0.394            \\ \hline
NMF (Clicks)                & 0.085            & 0.101            & 0.445             & 0.468            \\ \hline
NMF (CTR)                   & 0.102            & 0.103            & 0.371             & 0.420            \\ \hline
k-means+w2v (Clicks) & 0.084            & 0.100            & 0.455             & 0.437            \\ \hline
k-means+w2v (CTR)    & \textbf{0.125}   & \textbf{0.139}   & \textbf{0.467}    & \textbf{0.475}   \\ \hline
\end{tabular}
\end{center}
\end{table}

Table \ref{tbl:result_all} shows the results for all metrics in both offline experiments. 
According to these metrics and experiments, the proposed method of k-means+word2vec (CTR) demonstrates the best performance. 
In the {\it all experiment}, MAP@10 improved by 22.5\% compared to the highest baseline, NMF (CTR), and NDCG@10 improved by 38.6\% compared to same baseline. 
In the {\it user experiment}, MAP improved by 2.03\% compared to the highest baseline, NMF (clicks), and NDCG improved by 1.07\% compared to the highest baseline, a simple content-based method. 
Based on the above, the proposed method provides improvements over the baseline.
Since there were too many candidate articles in the {\it all experiment}, the overall performance was lower than that of the {\it user experiment}.
In the {\it all experiment}, the performances of the simple content-based
and clicks-based methods declined more rapidly than they did in CTR-based methods.
Simple content-based and click-based methods were overestimated in the {\it user experiment} because of the existing CTR-based system, as discussed earlier.

\begin{figure}[tp]
  \centering
  \includegraphics[width=0.99\linewidth]{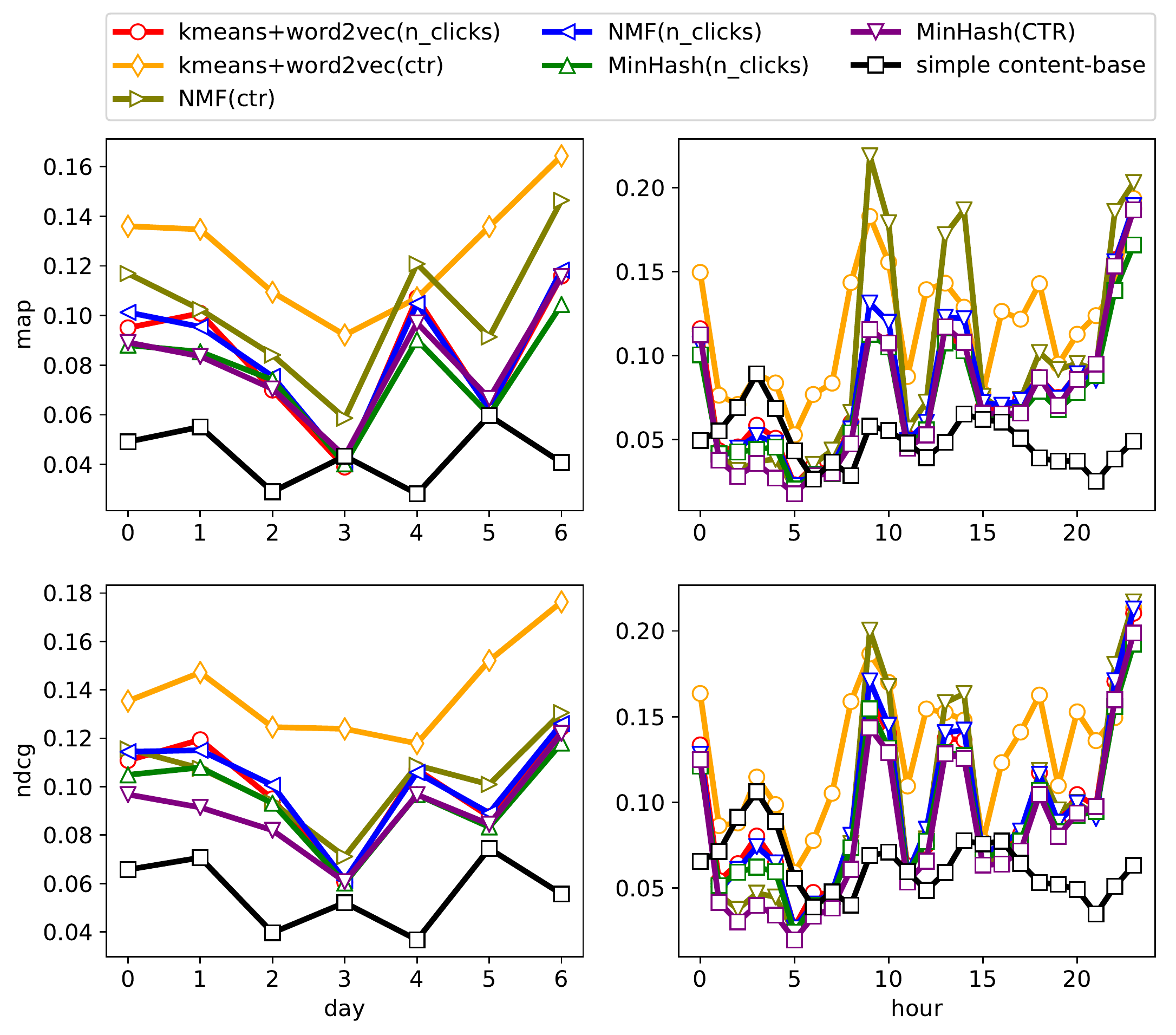}
    \caption{The {\it All Experiment} Daily and Hourly Results: On a daily basis, the proposed method is shown to be the best on every day but one. On an hourly basis, several baselines outperformed the proposed method in certain hours.}
    \label{fig:all_exp_result}
\end{figure}

\begin{figure}[tp]
  \centering
  \includegraphics[width=0.99\linewidth]{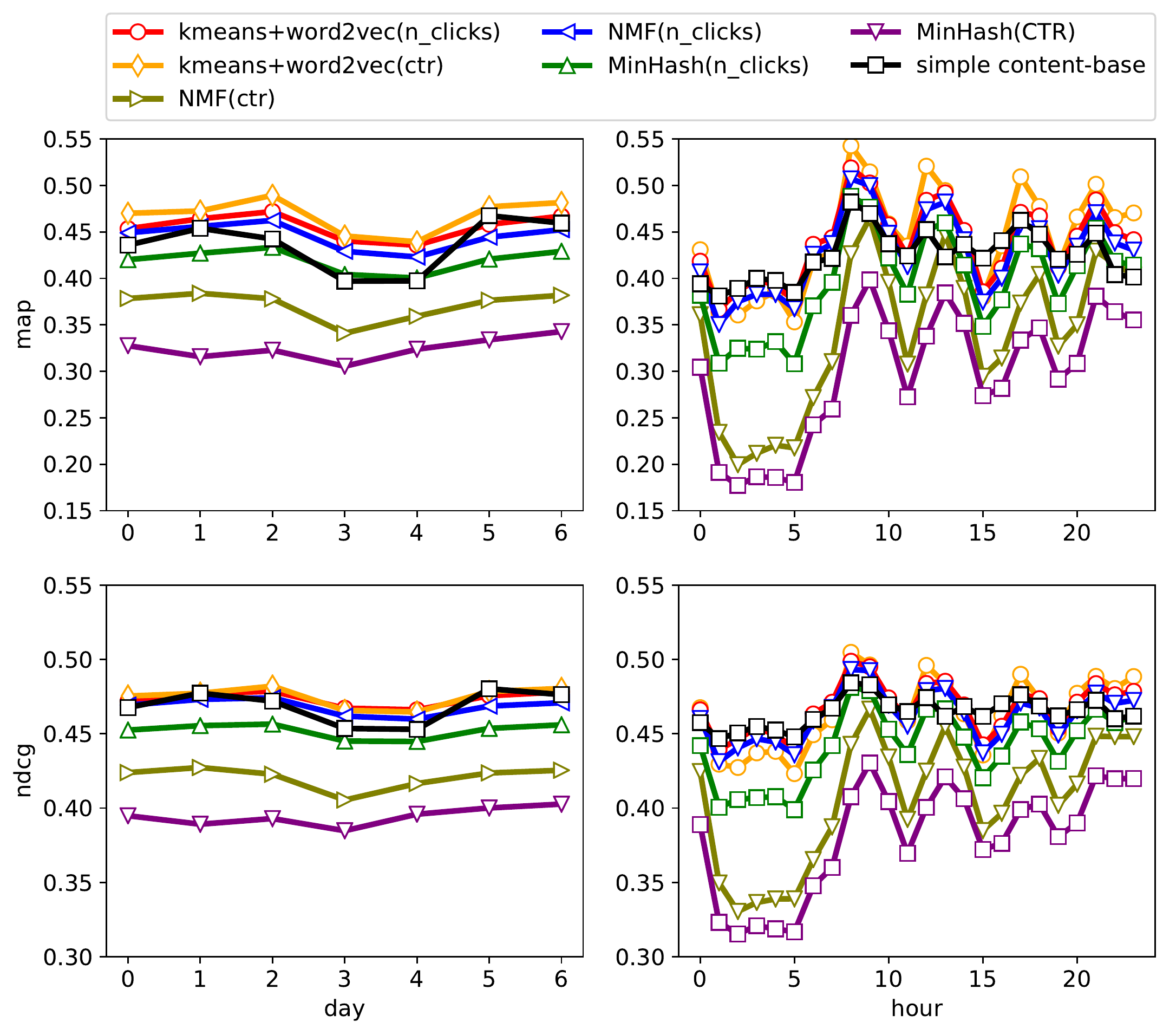}
    \caption{The {\it User Experiment} Daily and Hourly Results: On a daily basis, the proposed method is shown to be the best on every day. On an hourly basis, several baselines outperformed the proposed method in certain hours.}
    \label{fig:user_exp_result}
\end{figure}

Figure \ref{fig:all_exp_result} and Figure \ref{fig:user_exp_result} show the performances of these experiments on a daily and hourly basis. 
In the {\it all experiment}, the proposed method proved to be the best method among all days/hours except for one single day/hour.
Moreover, the same is true for the {\it user experiment}.

According to these results, our method performed in the offline experiments.
Our method is far superior to other baselines in speed, scalability, and updateability in the online environment. Thus, our method qualifies as a  suitable method for being deployed on million-scale production environments. 

\subsection{Online Experiments}

In our online experiments (A/B testing \cite{Ron}), users were divided evenly into groups, each using a different method, and we deployed the proposed and compared methods in NewsPass.
In the experimental setting, we measured three metrics: CTR, article clicks per session (Clicks/Sessions), and clicks per user per session (Click Users/Sessions). 
Although the ratio and the number of users tested are not described for a business reason,
this restriction does not impair the effectiveness of the result. 

We compared the following three methods: 
\begin{enumerate}
\renewcommand{\labelenumi}{(\roman{enumi})}
\item {\bf Control}: existing methods that employ user demographics and article CTR;
\item {\bf k-means+w2v+TDF}: the proposed method that uses a k-means clustering algorithm, word2vec, and the TDF in equation (\ref{tdf}); and
\item {\bf k-means+w2v+UTDF}: the proposed method that uses a k-means clustering algorithm, word2vec, and the UTDF in equation (\ref{utdf}).
\end{enumerate}
The thresholds of elapsed time were set at 1 hour for $T_{\it tdf}$ and 4 hours for $T_{\it utdf}$. 
These settings were decided empirically.

\begin{table}[htb]
  \begin{center}
    \caption{The Result of Online Experiment: These metrics are described as a ratio when the metric of {\it control} is set at $1.0$. These results show that our method outperforms the control and the UTDF outperforms the TDF.}
    \label{online-test}
    \begin{tabular}{l r r r}
      Method & CTR  & Clicks/Sessions & Click Users/Sessions\\ \hline
      Control & 1.0000 & 1.0000 & 1.0000 \\
      TDF & 1.0106 & 1.0332 & 1.0032 \\
      UTDF & \textbf{1.0318} & \textbf{1.0576} & \textbf{1.0075} \\  \hline 
    \end{tabular}
  \end{center}
\end{table}

The experimental results are shown in Table \ref{online-test}. 
These metrics are described as a ratio when the control metric is set at $1.0$.
We confirmed the effectiveness of this method by testing each metric through the Chi-squared test, but details are not described for a business purpose.
These results show that our method outperforms the control and that the UTDF outperforms the TDF.
As a result, the effectiveness of our proposed method was confirmed by both the online and offline experiments, 
and score decay based on a user's last access time was found to activate the user's behavior.
Moreover, we confirmed that our system worked stablly in our million-scale production environment.


\section{Related Works}\label{section:related_work}
Recommendation systems are an active research area and several information filtering algorithms have been proposed for a variety of items \cite{Pazzani, Su, Bobadilla}, such as movies \cite{Christakou}, music \cite{Van, Wang-Xinxi}, and applications \cite{Cheng}.
A variety of production services (e.g., Amazon, YouTube, and Netflix) have introduced recommendation systems to allow customers to make more effective use of their services \cite{Youtube, Netflix}.

This paper focused on news recommendation systems.
In news recommendation, the popularity of candidate news articles changes rapidly because the content value of news quickly decays; \cite{Zheng-Guanjie} reported that the average lifespan of breaking news is about 4.1 hours.
Moreover, users' interests also change rapidly compared to other kinds of items (e.g., movies, products, or restaurants) \cite{ozgobek2014survey}.
Therefore, we propose immediate news recommender systems.
Immediate means that changes in news trends and user interests are reflected in recommended news lists quickly.
Although some personalized online news recommendation systems have been proposed \cite{google, google2, Li-Lei,Zheng-Guanjie}, 
no studies have provided solutions to reflect and update these changes immediately.

In \cite{google, google2, Li-Lei}, the first user cluster was based on MinHash and pLSI clustering and was used to calculate cluster weights for each user.
Their systems recommend news based on the number of clicks and the CTR of news articles for each cluster.
When a user clicks on a news article, the number of clicks for that user cluster is updated, and then, their systems recommend the news article for the same cluster of users in real time.
The main differences between our method and existing methods are clustering and updating user vectors.
Our method clustered users based on k-means and word2vec-based vectors using user`s click history.
We expected these vectors to accurately reflect the preferences of users who have a brief click history.
Furthermore, when a user clicks on a news article, our method updates not only the number of clicks but also the user vectors.
By updating user vectors in real time, our method can capture changes in user preferences.

In \cite{Zheng-Guanjie}, news was recommended using deep reinforcement learning.
However, it is difficult to capture changes in user preferences using this method because users' features are only updated hourly or daily.

In \cite{okura}, a recurrent neural network was used to recommend news, with browsing histories as input sequences.
The authors proposed the content-based method to avoid suffering from the dynamic changes of news recommendations.
However, we proposed a collaborate filtering-based method because our offline experiments showed that this method achieves a better performance than content-based methods.
Instead of word2vec, distributed representation based on a recurrent neural network is applicable to our method.

\section{Conclusion}\label{section:conclusion}

In this paper, we examined algorithms and system architecture in immediate personalized news recommendation systems. 
Although many news recommendation systems have been proposed, our system focuses particularly on immediacy. 
Immediacy means changes in news trends and user interests are reflected in recommended news lists quickly. 
We expected that an immediate news recommendation system would be able to quickly recommend high-value news articles to users. 

To capture the change of article popularity immediately in a personalized way, our algorithms are based on collaborative filtering (CF) of user clusters and evaluates news articles through click-through rate (CTR) and decay scores using the time elapsed since the user`s last access. 
One of the challenges of CF is that it is difficult to evaluate new items properly.
The reason for using CTR for scoring articles is it aims to reflect current popularity.
To evaluate the freshness correctly in a personalized way, we calculated how new a news article is for each user using the user`s last access time.

To capture the change in users' interests immediately, we designed the user model to be updated as soon as they click on a news article.
New users` vector representations can be obtained immediately after they click on only one news article.

Moreover, we provided the architecture of our proposed system.
Although system architecture is important to deploy recommendation system on large-scale production environments, existing studies have not fully discussed system architecture.
Therefore, a major contribution of our research is that we reveal our system architecture to realize our algorithms and a configuration example implemented on top of AWS.

We evaluated the proposed system offline and online. 
The offline experiments were conducted using a real-world dataset from a commercial news delivery application. 
Results indicate that our user modeling and cluster-based CF via CTR method is effective. 
The online experiment was conducted through an A/B test on a real service. 
We confirmed that the effectiveness of our proposal method and the score decay based on a user's last access time improves user experiences.

Currently, the system is implemented and deployed on our services, Gunosy,\footnote{https://gunosy.com/} NewsPass, and LUCRA.\footnote{https://lucra.jp/}
The total number of users on these applications amounts to millions. 
We plan to collect further explicit feedback and user profiles to valuate news articles more profoundly in the future.

\begin{acks}
We would like to thank the Engineering Teams of our news delivery services, the Machine Learning Team, and the Data Management Platform Team for their contributions to this project.
\end{acks}

\bibliographystyle{ACM-Reference-Format}
\bibliography{ref}
\end{document}